\documentclass[aps, pra, superscriptaddress, reprint, twocolumn,]{revtex4-2}
 
\usepackage{graphicx,color}
\usepackage{amsmath,amssymb,amsfonts,amsthm,amscd,bm}
\usepackage{braket}
\usepackage{subcaption}

\graphicspath{{pictures/}}

\newcommand{\abs}[1]{\left| #1 \right|}
\newcommand{\norm}[1]{\left| #1 \right|^2}
\newcommand{\dd}{\mathrm{d}}
\newcommand{\comment}[1]{}

\begin{document}

\title{
Reflection and Transmission Amplitudes \\in a Digital Quantum Simulation
}
\date{\today}

\author{Giuseppe Mussardo}
\affiliation{SISSA and INFN, Sezione di Trieste, via Bonomea 265, I-34136, Trieste, Italy}

\author{Andrea Stampiggi}\email{astampig@sissa.it, corresponding author} 
\affiliation{SISSA and INFN, Sezione di Trieste, via Bonomea 265, I-34136, Trieste, Italy}

\author{Andrea Trombettoni}
\affiliation{Dipartimento di Fisica, Universit\'a di Trieste, Strada Costiera 11, I-34151 Trieste, Italy and\\
SISSA and INFN, Sezione di Trieste, Via Bonomea 265, I-34136, Trieste, Italy}

\begin{abstract}
  In this paper we show how to measure in the setting of digital quantum simulations the reflection and transmission amplitudes of the one-dimensional scattering of a particle with a short-ranged potential. The main feature of the protocol is the coupling between the particle and an ancillary spin-1/2 degree of freedom. This allows us to reconstruct tomographically the scattering amplitudes, which are in general complex numbers, from the readout of one qubit. Applications of our results are discussed.
\end{abstract}

\maketitle

\section{Introduction}\label{s_intro}
One of the most striking differences between classical and quantum mechanics is that a particle -- or generally a many-particle system -- can tunnel through a potential barrier. As such, even if classically a particle with an energy lower than the barrier can only be reflected by a localized potential, there is a nonzero probability of finding its quantum analogue also transmitted \cite{roy1986quantum}. This remarkable feature has both conceptual implications and practical applications. On the one hand it permits the implementation of superconducting quantum bits \cite{makhlin01} and quantum devices based on tunneling phenomena \cite{barone1982physics}: these are nowadays routinely used for quantum sensing and several other applications \cite{wolf2017josephson}. On the other, the determination of scattering amplitudes is a fundamental problem of quantum field theory \cite{badger2023}, not only for estimating cross-sections, but also for carrying on the evaluation of several other observables, from both a perturbative \cite{peskin2018introduction, colemanbook} and non-perturbative point of view \cite{mussardo2020statistical}.

The central role played by scattering amplitudes in this wide range of applications and the potentialities of quantum computation  have been stimulating communities from high-energy physics \cite{dimeglio2023, 2023bauer} to lattice gauge theory \cite{banuls2020simulating, Funcke2023}. There is hope for an interesting synergy: the possibility to extract quantities related to scattering from quantum simulations \cite{georgescu2014, Halimeh2023}. In an analog quantum simulation \cite{Bloch2012,Gross2017}, the goal is to use a physical platform, such as ultracold atoms \cite{cornish2024quantum} or ion traps \cite{blatt2012quantum}, to emulate the physical system of interest and determine quantities, otherwise difficult to find with other approaches. On the other hand, the platform of a digital quantum simulation is a quantum computer, which can be in principle programmed with a universal language of computation \cite{r_2010nielsen_b_qc_qi}. Through the mapping of the original system to a system of interacting spin-1/2 (qubits), it is possible to obtain estimates for the physical quantities of the original system from the readout of the qubits. This latter approach has been used in \cite{jordanleepreskill} to provide a framework to determine the scattering amplitudes of an interacting scalar field theory with quartic coupling. More recently, an algorithm for theories with bound states has been proposed \cite{2024turco}. On the other hand, there are other approaches aimed at determining scattering amplitudes through different strategies and methods, e.g. from correlation functions \cite{guopenggasparian-scattering}. Our contribution to this expanding field is an algorithm of quantum computation, including state preparation, time evolution and readout process, for the determination of scattering properties of the simplest possible model of quantum mechanics: a single non-relativistic particle scattering with a localized potential.

More specifically, in this paper we focus on the digital quantum simulation of the quantum tunneling of a one-dimensional particle and on measurements of reflection ($\mathcal{R}$) and transmission ($\mathcal{T}$) amplitudes for an arbitrary localised barrier. As well known, the one-dimensional scattering of a quantum particle features several remarkable qualities and can be of guideline for the analysis of more complicated scenarios. First, it is a textbook example \cite{r_1998merzbacher_b_qm,CohenTannoudji} in which one can solve exactly the scattering problem in many instances (e.g., for piecewise potentials). Moreover, for the case of parity symmetric potentials, it is possible to relate these scattering coefficients to the phase shifts one normally considers in the scattering theory in higher dimensions -- see \cite{lipkin1973quantum} and Appendix~\ref{a_smatrix} for details. Finally, one can straightforwardly put in connection the scattering amplitudes for the non-relativistic particle with the $S$-matrix of $1+1$ interacting field theories \cite{mussardo2020statistical} and in turn one can also determine the $S$-matrix from $1+1$ conformal field theories defined on the cylinder 
\cite{Lassig1996}. As we will argue, the digital quantum simulation of the one-dimensional scattering is already equipped with the structure of a more general protocol, which can be applied to simulations of particle scattering in higher dimensions. 

It is useful to underline that reflection and transmission coefficients do not correspond to observables, i.e. they are not eigenvalues of Hermitean operators. However, they can still be determined by repeating a physical process -- in this case the scattering of a particle with a localised potential -- a statistically significant amount of times, as one would ideally do in an experiment.

The key idea we will pursue here is firstly to encode the information about the particle scattering in the components of an ancillary spin-1/2 degree of freedom and secondly to perform a quantum tomography of this state. Indeed, a pure spin-1/2 state can be fully reconstructed by the averages of the measurements of the three components of the spin. Interestingly enough, as discussed below, this protocol can be also implemented in a very natural way as a digital quantum simulation, i.e. as an algorithm of quantum computation, in which the particle state involved in the scattering is encoded in many interacting qubits.  

Another novel result of this work concerns the readout process of the simulation. Indeed standard treatments of this problem \cite{r_1998zalka_qc_simulation, r_2008benenti_qc_oneparticle_scattering, r_2010nielsen_b_qc_qi} usually limit themselves to reconstructing the probability density of the particle from the measurement of all qubits. More recently, the simulation of one-dimensional dynamics in a well of infinite height has been studied \cite{ostrowski2024}, for which no tunneling effect can happen and therefore cannot be compared to our work, which instead seeks to study the scattering properties in presence of localized potentials. Moreover, we are not interested here in the whole wave-function, but only in the values of scattering amplitudes. Therefore one expects that the measurement protocol should admit a more optimized version, i.e. the possibility to make less iterations of the algorithm in order to obtain meaningful answers. 
Indeed, this is true since it turns out that reflection and transmission coefficients can be obtained from the values of only one of the qubits, namely the one associated to the sign of the physical momentum of the particle.

\begin{figure}
    \centering
    \includegraphics[scale = 0.25]{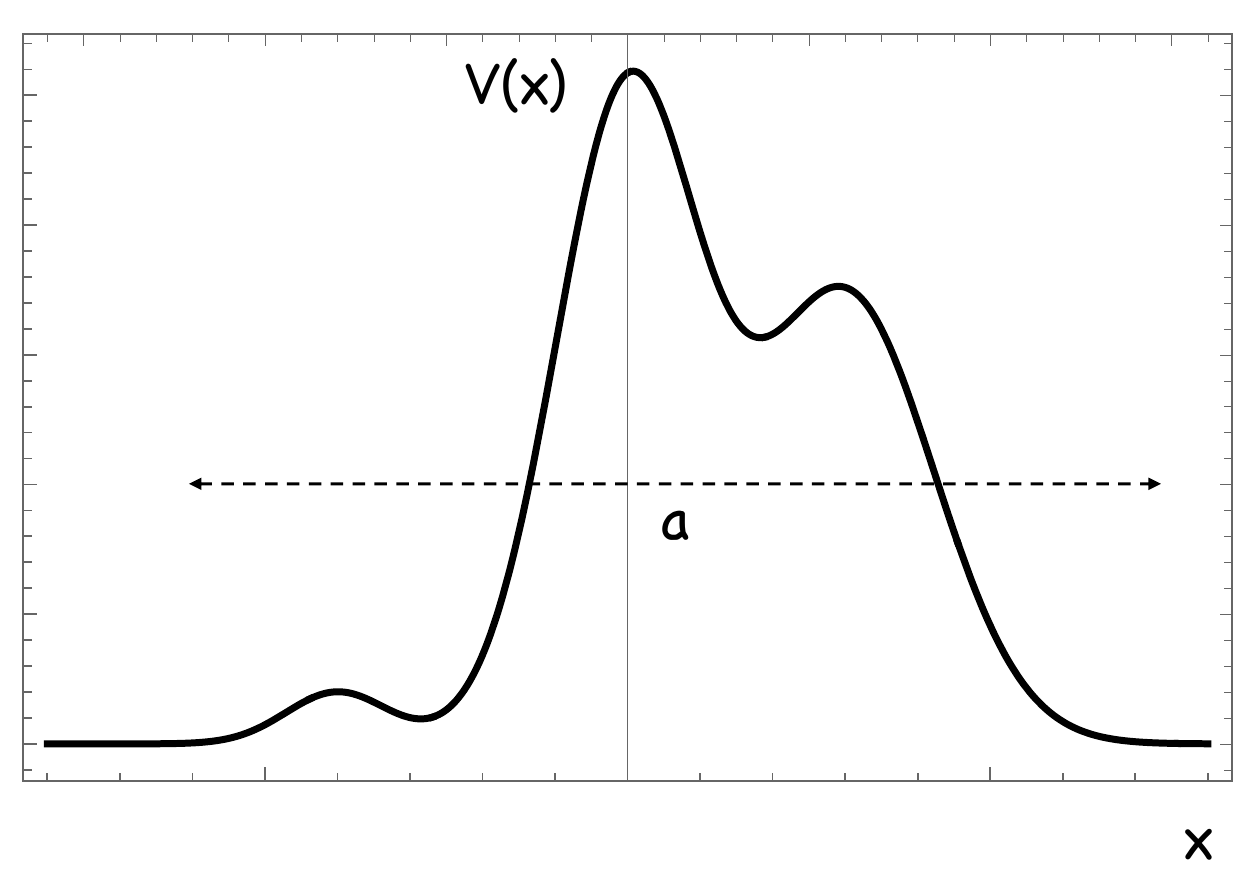}
    \caption{A generic localised potential barrier $V(x)$, bounded from below and nonzero only in a region of length $a$.}
    \label{barrierrr}
\end{figure}

In the following we consider the one-dimensional scattering of a non-relativistic particle of mass $m$ and momentum $p=\hbar\kappa$ on a potential barrier $V(x)$ centered around the origin. The information about this scattering is encoded in the initial and conserved energy of the particle $E = \hbar^2 \kappa^2 /2m$ and in the shape of the potential barrier. We assume the potential barrier $V(x)$ to be a non-negative function, with the only important qualitative property to be localised and different from zero only in a region of length $a$ around the origin, see Fig.~\ref{barrierrr}. Apart from these requirements, we assume that the shape of $V(x)$ is arbitrary. Our protocol, in absence of known analytic expressions for the reflection and transmission amplitudes of an arbitrary potential, sets up a digital quantum scheme which allows us to compute them.

The finite range $a$ of the potential $V(x)$ implies that any eigenfunction of the 
Schr\"{o}dinger equation is asymptotically a superposition of plane waves. For an incident particle from the left, we define as usual \cite{r_1998merzbacher_b_qm,CohenTannoudji} the reflection and transmission amplitudes according to the asymptotic behavior of the wave-function: 
\begin{equation}\label{eq_asymptotics_eigenfunctions}
\varphi_\kappa(x) \,=\,  \begin{cases}
e^{i\kappa x} + \mathcal{R}_\kappa \, e^{-i\kappa x}, & x\ll -a/2,\\
\mathcal{T}_\kappa \,e^{i \kappa x}, & x\gg a/2.
\end{cases}
\end{equation}
For an incident particle from the right, analogous results hold if one considers a solution of opposite momentum. They coincide if the potential is a symmetric function with respect to the origin.

The conservation of probability currents, i.e. the unitarity condition of the scattering process, implies \begin{equation}\label{eq_unitarity}
\norm{\mathcal{R}_\kappa} + \norm{ \mathcal{T}_\kappa } = 1. 
\end{equation}
In view of this equality, we can regard $\mathcal{R}_\kappa$ and $\mathcal{T}_\kappa$ as the --generically complex-- components of a spin-1/2 degree of freedom: 
\begin{equation}\label{eq_spin_observation}
\ket{\varphi} = \mathcal{R}_\kappa \ket{-_{\kappa}} + \mathcal{T}_\kappa \ket{+_{\kappa}}.
\end{equation}
This turns out to be a crucial observation for the measurement of reflection and transmission amplitudes and in the following we will elaborate on it.

The content of the paper is organised as follows. We first recall in Sec.~\ref{s_spin} how to reconstruct the amplitudes of a spin-$1/2$ through measurements of polarization. Then we present in Sec.~\ref{s_protocol} a protocol to encode reflection and transmission amplitudes into an ancillary spin-$1/2$. A natural implementation of this protocol as a digital quantum simulation algorithm is discussed in Sec.~\ref{s_simulation}. Our conclusive remarks are gathered in Sec.~\ref{s_conclusions}. The paper has also two appendices. The first one, Appendix~\ref{a_smatrix}, frames reflection and transmission amplitudes within the general scattering theory formalism \cite{r_1998merzbacher_b_qm, lipkin1973quantum}. In this appendix we also show how to compute them numerically through the sampling of the potential with $\delta$-functions. In Appendix~\ref{a_kinetic} we discuss the implementation of the gates useful in the quantum simulation.

\section{Measurement of Spin-$1/2$ Amplitudes}\label{s_spin}
\begin{figure}
\centering
\includegraphics[scale = 1]{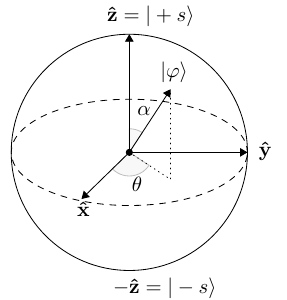}
\caption{Pictorial representation of the state $|\varphi\rangle$ on the Bloch sphere. By repeated measurements of the spin polarization, see Eq.~\eqref{eq_spin_tomography}, it is possible to 
fix the angles $\theta$ and $\alpha$ to arbitrary precision and therefore to reconstruct the spin state
$|\varphi\rangle$.}
\label{f_bloch}
\end{figure}

A general spin-1/2 state can be parameterized on the Bloch sphere (Fig.~\ref{f_bloch}) as
\begin{equation}\label{eq_final_state_angles}
|\varphi \rangle = \sin(\alpha /2) |-_{s}\rangle + \cos(\alpha/2) e^{i\theta}|+_{s}\rangle.
\end{equation}
On this state, the expectation values of spin along the three axes can be reconstructed by repeated measurements of the three
components of the spin \cite{r_2018benenti_b_qc}. If $\mathcal{P}$ are their probabilities, e.g. $\mathcal{P}_z^+$ for measuring $+1$ on $\widehat{\sigma}_z$ and so on, they satisfy
\begin{equation}\label{eq_spin_tomography}
\begin{aligned}
\mathcal{P}_x^+ - \mathcal{P}_x^- &= \sin \alpha \cos \theta, \\
\mathcal{P}_y^- - \mathcal{P}_y^+ &= \sin \alpha \sin \theta
\quad \text{and} \\
\mathcal{P}_z^+ - \mathcal{P}_z^- &= \cos \alpha.
\end{aligned}
\end{equation}
These equations show that it is possible to perform a full tomography of an arbitrary spin-1/2 state. 

\section{Protocol}\label{s_protocol}
The outcome of the scattering of a quantum particle with a localized potential consists of a nonzero probability of finding its wave-function both transmitted and reflected. For definiteness, consider the potential $V(x)$ to be non-zero only in a region $-a/2 \leq x \leq a/2$ and the initial particle to be of positive momentum $p = \hbar \kappa$, well-localized around a point $-x^{0} \ll -a/2$. Such approximation holds if $x^{0}$ is much greater than the de Broglie wavelength $\lambda_{\text{dB}} = h/p$. In this section we provide a description of the ideal protocol, discussing the physical interpretation and the details of the digital quantum simulation in the next section.

We prepare the initial state 
\begin{equation}\label{eq_initial_state}
\ket{\phi_0} = \ket{+_{\kappa}}\otimes \ket{+_s}
\end{equation}
with its ancillary spin-$1/2$ $\textbf{S}$ along the positive $z$-direction -- if the particle is, for instance, an electron, the ancilla coincides with the actual electron spin. This state can be prepared at asymptotic times by a series of measurements, which finally filter the desired state. First, we imagine that there is a device capable of producing a particle with positive and negative momentum with arbitrary spin. Since the whole point of this protocol is not to measure the momentum, we can turn on an interaction 
\begin{equation}\label{eq_hamiltonian_momentum_spin}
\widehat{H}_{\textbf{P}, \textbf{S}} = - g \left(\text{sign}(\widehat{P})\widehat{\sigma}_z-1\right),
\end{equation}
for $g>0$ and then measure the energy of the state, selecting only the ground state of this Hamiltonian. After this step, we measure the spin of the particle and select $\ket{+_s}$. We then switch on the potential $V(x)$, a procedure that turns out to be useful also in the study of scattering of properties of the non-linear Sch\"rodinger equation where the reflection and transmission coefficients cannot be defined as above due to the non-linearity of the equation \cite{Paris-Mandoki2017}. 

At time $t=0$, we then let the particle scatter with the potential, keeping in mind that the interaction with the potential barrier does not involve the spin degree of freedom
\begin{equation}\label{eq_shrodinger_hamiltonian}
\widehat{H}_{\textbf{P}} \,=\, \frac{\widehat{P}^2}{2m}+V(\widehat{X}).
\end{equation}
Estimating the proper time scales of tunneling is an issue under debate -- see e.g. \cite{landauerBarrierInteractionTime1994} for a review. For the purposes of the present discussion, only a time-scale able to gauge whether if the outgoing wave is asymptotically free is needed. A conservative estimate is given by the semi-classical motion of the peak of the packet, which defines a time-scale $t_s \approx 2x_0 m/p$, corresponding to the state
\begin{equation}\label{eq_state_scattering}
\ket{\phi_{t_s}} = \left(\mathcal{R}_{\kappa} \ket{-_{\kappa}} + \mathcal{T}_{\kappa} \ket{+_{\kappa}} \right) \otimes \ket{+_s}.
\end{equation}

\begin{figure}
    \centering
    \includegraphics[scale = 1]{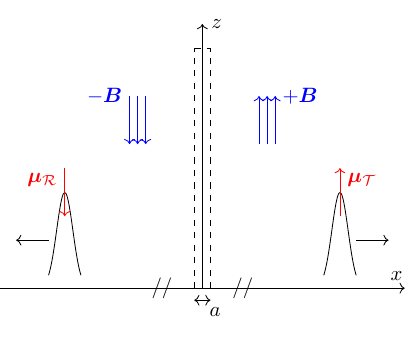}
    \caption{In the semi-classical picture, the magnetic moment of the particle is aligned with the direction of the reflected and transmitted waves by switching on a magnetic field in opposite directions to the left and right of the potential. The interaction is the usual spin-orbit one: $H = -\bm{\mu}\cdot\bm{B}$, where $\bm{\mu}$ is the magnetic moment and $\bm{B} = B\text{sign}(x)\bm{z}$ the magnetic field.}
    \label{f_spin_momentum}
\end{figure}

In a semi-classical way to describe such a scattering, after the time interval $t_s$ we may regard the state of the system as made up of two outgoing waves propagating with opposite velocities: the reflected wave to the left of the potential and the transmitted one to its right. Crucially, after the time scale $t_s$ these waves are minimally overlapping in the potential region and therefore they are propagating freely. In order to align spin and momentum in this setting one can switch on a magnetic field having opposite direction to the left and right of the potential, see Fig. \ref{f_spin_momentum}.

On a quantum mechanical level, we can obtain the same result by means of a unitary evolution. Since in the region to the left of the potential it is most probable to find the reflected packet, we apply the interaction
\begin{equation}
\widehat{H}_{\textbf{S}} = \frac{g'}{2}\widehat{\sigma}_x, \quad \text{for }x<0
\end{equation}
and for a time $\delta t= \pi/g'$. To the region to the right of the potential, no interaction between the ancillary spin and momenta is applied. The unitary operation obtained through exponentiation is equivalent in the quantum computing language to the application of a controlled operation  -- in this case the C-NOT -- and allows us to find the desired final state, i.e. one in which the information about the scattering is encoded into the ancillary spin degree of freedom:
\begin{equation}\label{eq_gedanken_momentum_spin}
\ket{\phi^> } = -i \mathcal{R}_{\kappa} \ket{-_{\kappa}, -_{s} } + \mathcal{T}_{\kappa} \ket{+_{\kappa}, +_{s}}.
\end{equation}

Then, by performing measurements of spin as in Sec.~\ref{s_spin}, we can extract not only the values of the modulus of $\mathcal{R}_{\kappa}$ and $\mathcal{T}_{\kappa}$, but also of their relative phase.

\section{Digital Quantum Simulation}\label{s_simulation}
The first step to set up the digital quantum simulation of particle scattering is to define the proper dimensionless  quantities for numerical computation. To this aim is convenient to consider the usual Schr\"{o}dinger equation for a particle of mass $m$, potential $V(x)$ and initial wave-function $\psi(x)$
\begin{equation}\label{eq_1D_shrodinger}
i\hbar \partial_t \psi(x,t) = \left[-\frac{\hbar^2}{2m}\partial_x^2 + V(x) \right]\psi(x,t).
\end{equation}
In any computation only a finite amount of points can be sampled. Hence, let $L$ be the interval range which contains the relevant physics of the problem, with $L \gg a$ and chosen as follows: given a threshold $\varepsilon$, we choose $L$ such that 
\begin{equation}
\Delta_L:= \int_{-\infty}^\infty \dd x\, \norm{\psi(x)} - \int_{-\frac{L}{2}}^{\frac{L}{2}} \dd x\, \norm{\psi(x)} \leq \varepsilon.
\end{equation}
The wave-function $\psi(x)$ can then be safely truncated without compromising the physical properties of the system:
\begin{equation}\label{eq_wavefunction_truncation}
\psi_\varepsilon (x):= \begin{cases}
\frac{1}{\sqrt{1-\varepsilon}}\psi(x), &\text{ if } -L/2\leq x \leq L/2, \\
0, &\text{otherwise}.
\end{cases}
\end{equation}
At this point, we introduce length and time scales and express all quantities in dimensionless units. For convenience, we resize the system in the unit interval $\xi = x/L \in \left[-1/2, 1/2\right]$ and define the dimensionless time $\tau = t/T$, where $T$ is some unit of time. If we are simulating a wave packet from the very beginning, we can choose $T$ as the time $t_s$ of Sec.~\ref{s_protocol} --see also Sec.~\ref{s_readout}.

The (dimensionless) wave-function is then $f(\xi) = \sqrt{L}\psi(x)$, the potential $u(\xi) = \frac{2m L^2}{\hbar^2} V(x)$ and the mass is expressed in terms of the parameter $\gamma = \frac{\hbar}{2m}\frac{T}{L^2}$. With such definitions, the Schr\"{o}dinger equation reads
\begin{equation}\label{eq_adimensional_shrodinger}
i \partial_\tau f(\xi,\tau) = \gamma \left[-\partial_\xi^2 + u(\xi)\right]f(\xi,\tau).
\end{equation} 

\begin{figure}
\centering
\includegraphics[scale = 1]{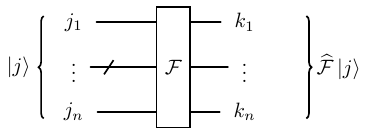}
\caption{The basis $\{\ket{j}\}$ is related to its conjugate $\{\ket{k}\}$ by the 
qFT $\mathcal{F}$ Eq.~\eqref{eq_qFT}. From the circuit it is evident that while the two basis have distinct interpretations, they are equivalent as far as the implementation of unitaries is concerned.}
\label{f_j_to_k}
\end{figure}

The digital quantum simulation involves a register of $n$ qubits $\{ \ket{j_i} \}$, each associated to binary variables $j_i = \{ 0,1\}$. As usual we choose to write any number $j = \{ 0, 1, \ldots, 2^{n}-1\}$ as
\begin{equation}\label{eq_convention_bits}
j = \overline{j_1 j_2 \ldots j_n} = j_1 2^{n-1} + j_2 2^{n-2}+\ldots+ j_n.
\end{equation} 
This notation also applies to any state $\ket{j}$ of the simulation. The original basis $\{ \ket{j}\}$ admits a conjugate basis $\{ \ket{k}\}$, defined through the ``quantum Fourier Transform'' (qFT) $\mathcal{F}$
\begin{equation}\label{eq_qFT}
\widehat{\mathcal{F}}\ket{j} = \frac{1}{2^{n/2}}\sum_{k=0}^{2^n-1} e^{-2\pi i jk/2^{n}} \ket{k},
\end{equation}
which requires $O(n^2)$ gates to implement \cite{r_2010nielsen_b_qc_qi}. Crucially, the states in the conjugate basis are associated to the integers
\begin{equation}\label{eq_convention_k}
k = \overline{k_1 k_2 \ldots k_n} = k_1 2^{n-1} + k_2 2^{n-2}+\ldots+ k_n,
\end{equation} 
which are in the same number as the $j$'s (see Fig~\ref{f_j_to_k}).

Since we ultimately want also to connect the $k$'s to physical momenta, we discretize position as follows:
\begin{equation}\label{eq_digitalization_space}
\xi_j = j 2^{-n} - \frac{1}{2}.
\end{equation}
The sampled wave-function has points 
\begin{equation}\label{eq_digitalization_wavefunction}
f_j = \frac{1}{\sqrt{\mathcal{N}}} f(\xi_j), \quad \mathcal{N} = \sum_{j=0}^{2^n-1} |f(\xi_j)|^2 \delta_\xi,
\end{equation}
and it is immediately seen from the application of the Euler-McLaurin formula that the error in such approximation is $\abs{1-\mathcal{N}}= O(2^{-n})$. Analogously, we sample the potential as $u_j = u\left(\xi_j\right)$.

\subsection*{Overview of the Algorithm}\label{s_overview}
In the following, the algorithm is presented in a form suitable to simulation: all $n$-qubit operations are written in terms of single-qubit and CNOT operations. In Sec.~\ref{s_preparation} is discussed how to prepare the initial state. In Sec.~\ref{s_evolution} the time evolution protocol is presented, with its algorithmic step given by Eq.~\eqref{eq_time_evo_circuit}. Finally, the readout to extract reflection and transmission amplitudes is discussed in Sec.~\ref{s_readout}.

\subsection{State Preparation}\label{s_preparation}
\begin{figure}
    \centering
    \includegraphics[scale = 0.6]{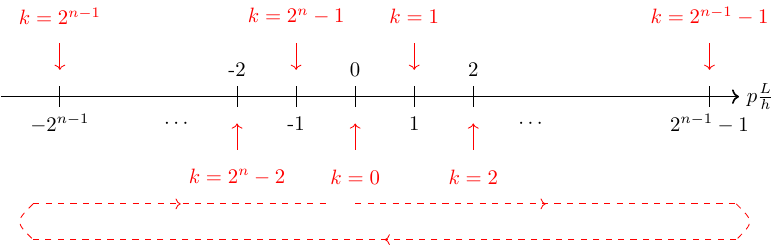}
    \caption{Relation between the physical momenta and the positive integers $k$, according to Eq.~\eqref{eq_momentum_k}.}
    \label{f_momenta}
\end{figure}

Our simulation starts with the following initial state
\begin{equation}\label{eq_initial_state}
\ket{f_0}= \frac{1}{\sqrt{\mathcal{N}}} \sum_{j=0}^{2^n-1} f_j \ket{j}.
\end{equation}
Before discussing how to initialize this state from $\ket{0}$, let us analyze the relation between the protocol of discretization and the physical variables, i.e. position and momentum. First of all, let's notice that the qFT on the initial state induces a discrete Fourier Transform in the space of $k$'s
\begin{equation}
\widehat{\mathcal{F}}\ket{f_0} = \ket{F_0} = \frac{1}{\sqrt{\mathcal{N}}} \sum_{k=0}^{2^n -1} F_k \ket{k}, 
\end{equation}
where indeed 
\begin{equation}\label{eq_dFT_quantum_computing}
F_k = \frac{1}{2^{n/2}}\sum_{j=0}^{2^n -1} e^{-2\pi i jk 2^{-n}} f_{j}.
\end{equation}
On the other hand, physical momenta are defined in continuous space. Sampling the wave-function into $N$ points $f(J/N)$, where $J = \{ -N/2, \ldots, N/2-1\}$ makes the momenta $K= p L/h$ discrete in the range in $K= \{ -N/2, \ldots, N/2-1\}$. The wave-function in momentum space reads
\begin{equation}\label{eq_dFT_momentum}
\tilde{F}_K = \frac{1}{\sqrt{N}}\sum_{J= - N/2}^{N/2 - 1} e^{-2\pi i K J /N} f(J/N).
\end{equation}
Connecting it with Eq.~\eqref{eq_dFT_quantum_computing}, we obtain
\begin{equation}\label{eq_momentum_wavefunction_k}
\tilde{F}_K e^{-i\pi K}= \begin{cases}
F_{K+2^{n}}, &\text{ if } K = \lbrace -2^{n-1}, \ldots, -1\rbrace,\\
F_K, &\text{ if } K = \lbrace 0, \ldots, 2^{n-1}-1\rbrace.
\end{cases}
\end{equation}
The last equation gives rise to a mapping between the discretization $k$ and the physical momenta $K$:
\begin{equation} \label{eq_momentum_k}
K = \begin{cases}
k , &\text{ if }k = \lbrace 0, \ldots, 2^{n-1}-1\rbrace,\\
k-2^{n}, &\text{ if }k = \lbrace 2^{n-1}, \ldots, 2^{n}-1\rbrace.
\end{cases}
\end{equation}
A visual interpretation of the latter equation is given in Fig.~\ref{f_momenta}. Notice also that the bit $k_1$ is associated to the sign of momentum: positive if $k_1=0$ and negative if $k_1=1$.

In the context of this digital quantum simulation, we have then  successfully associated $j$ to the position and $k$ to the momentum. The initialization of the initial state Eq.~\eqref{eq_initial_state} is complete if we provide a unitary transformation such that $\widehat{\mathcal{U}}_{\text{in}}\ket{0} = \ket{f_0}$. Unfortunately, if the input is the most generic sampled wave-function, the number of gates scales necessarily as $O(2^n)$. The reason of this comes from the generic statement that a unitary $U(2^n)$ can be decomposed into a string of C-NOTs and single-qubit gates \cite{r_1995barenco_qc_elementary_gates}. In the case at hand up to $(2^n-1)$ $SU(2)$ matrices of the form
\begin{equation}
U(f_j) = 
\begin{bmatrix}
\sqrt{1-|f_j|^2} & - f_j^* \\
f_j & \sqrt{1-|f_j|^2}
\end{bmatrix}
\end{equation}
must be implemented, if no special requirements about $f(\xi)$ are made. They initialize Eq.~\eqref{eq_initial_state} from $\ket{0}$ up to global phase, corresponding to that of one of the remaining point of the wave-function which is fixed by the normalization of the column vector.

However, a physically reasonable initial wave-function should be localized in an interval far away from the boundary and the potential, in order to minimize finite-size effects and to represent an asymptotic free particle, as in Sec~\ref{s_protocol}. In this case, less gates are required. Also, the more the packet is peaked, the less is the contribution coming from the tails; so that the wave-function can be easily truncated on a number of points which is much smaller than $2^n$.

Let us pause here to comment more on this point. Often, one starts from some momentum wave-functions $\tilde{F}_K$, which are related to the $F_k$'s through Eq. \eqref{eq_momentum_wavefunction_k}. The $F_k$'s will therefore contain information about the peak in position space, which we assume to be to the left of the potential, and the width of the distribution $\sigma_p$ -- as one may get convinced by studying the Gaussian case. If the peak velocity is $p^{0}>0$ and the distribution sufficiently narrow, the state with highest probability is $\ket{k^{0}}$ with $k_1=0$, according to Eq. \eqref{eq_momentum_k}. Not sensible to the computation are the values of the $F_k$ which fall below the machine precision; these will be interpreted as $0$'s. Reasonably, the packet will only have contributions from right-moving waves, so it is fair to assume that states with nonzero probability are found within the interval containing a minimum $k^{-}$ and a maximum $k^{+}$, both positive with $k_1=0$. Therefore the range $(k^{+}-k^{-})$ will not cover all the $n$ bits, but in practice a much smaller number, say $\ell_{-}$ for the range to the left of $k^{0}$ and $\ell_{+}$ to its right, such that $\ell_{\pm} \ll n$. Therefore, much less gates are required to initialize the wave-function of this problem, in momenta space. Then, if we want to obtain the wave-function in position space, it is only necessary to apply a qFT to the state.

As an example, let us consider the simplest case, which is also useful to fix the concept of the line of reasoning, i.e. that of a wave of momentum $p^{0}>0$ and peak position $x^{0}<0$ -- however delocalized in space. In continuous space, we can write the initial state as
\begin{equation}
\ket{\psi} = e^{2\pi i p^{0} x^{0}/ h} \ket{p} = \int_{-\infty}^{\infty}\dd x\ e^{2\pi i p^{0} (x-x^{0})/h} \ket{x}.
\end{equation}
If we were to implement naively the wave-function directly in position space, at most $2^{n}-1$ sampled points would be required, while in momentum space we only need to implement
\begin{equation}
\ket{0} \to  e^{2\pi i k^{0} j^{0}} \ket{k^{0}},
\end{equation}
which requires one phase shift and a SWAP operation. Actually, in Appendix~\ref{a_kinetic} we show that for the case of such transformations, which add a phase linear in $j$, we can work directly in position space and only $n$ phase-shift gates are required.

\subsection{Time Evolution}\label{s_evolution}
Given the initial state Eq.~\eqref{eq_initial_state}, the time evolution with respect to the Hamiltonian $\widehat{H} = \gamma [(2\pi)^2\widehat{K}^2 + u(\widehat{\xi})]$ can be approximated with the Trotter-Suzuki formula with step $\delta_\tau = \tau/N_\tau$ 
\cite{r_1996lloyd_qc_simulation} 
\begin{equation}
\widehat{\mathcal{U}}_\tau = e^{-i \widehat{H} \tau} = \left[\widehat{\mathcal{K}}_{\delta_\tau}\widehat{\mathcal{V}}_{\delta_\tau}\right]^{N_\tau} + O(\delta_\tau),
\end{equation}
where $\widehat{\mathcal{K}}_{\delta_\tau} = e^{-i \gamma (2\pi)^2 \hat{K}^2 \delta_\tau}$ and $\widehat{\mathcal{V}}_{\delta_\tau} = e^{-i \gamma u(\hat{\xi}) \delta_\tau}$. The potential and the kinetic operators are separately diagonal in the basis of position and momentum, respectively. In the digital simulation, one can switch between them by applying the qFT. As such, the circuit approximating $\widehat{\mathcal{U}}_\tau$ up to $O(\delta_\tau)$ is
\begin{equation}\label{eq_time_evo_circuit}
\includegraphics[scale=1]{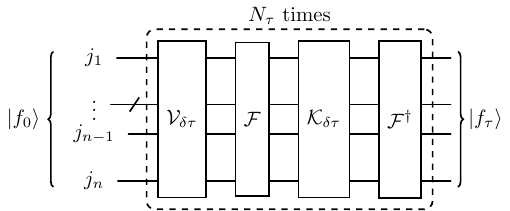}
\end{equation}

\begin{figure}
\centering
\includegraphics[scale = 1.]{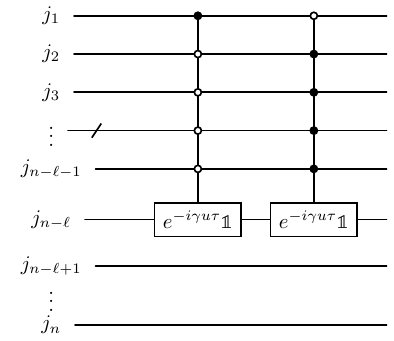}
\caption{Circuit performing the time evolution for a potential barrier of height $u$, centered in $j=2^{n-1}$ and of range $2^{2\ell}$, as in Eq.~\eqref{eq_potential_barrier}.}
\label{f_potential_barrier}
\end{figure}

In Appendix~\ref{a_kinetic} it is shown how $\widehat{\mathcal{K}}_{\delta_\tau}$ is realized in terms of $O(n^2)$ gates. As far as the potential is concerned, in the worst case scenario $2^{n-1}$ different controlled phase-shift are required, for the problem is equivalent to the implementation of a unitary transformation $\widehat{G}$ which assigns a phase $g(j)$ to each $\ket{j}$
\begin{equation}\label{eq_unitary_phase_shift}
\ket{\psi}  = \sum_{j=0}^{2^n-1} c_j \ket{j} \to \widehat{G}\ket{\psi}=\sum_{j=0}^{2^n-1} c_j e^{ig(j)} \ket{j}.
\end{equation} 
In general, $2^{n-1}$ controlled phase-shift operations of the form
\begin{equation}
G_m = \begin{bmatrix}
e^{i g(2m-2)} & 0\\
0 & e^{i g(2m-1)}
\end{bmatrix},
\end{equation}
for $m =\{ 1, 2, \ldots, 2^{n-1}\}$, must be performed. However, in the case of localized potentials the gate requirement can be significantly downsized. Assuming them to be centered around the origin $j=2^{n-1}$, we require them to occupy at most a range $2^{2\ell}$, where $\ell\ll n$ is the number of bits in which the potential is effectively nonzero. This condition, together with the assumption for the preparation of the initial state, are consistent with the requirement of an asymptotic free particle. Assuming in generality that the potential is not symmetric, we define two samplings, corresponding to the regions to the left and right of the origin:
\begin{equation}
u_j = \begin{cases}
u^{(>)}_j & \text{ if } j=\lbrace 2^{n-1}, 2^{n-1+\ell}-1\rbrace,\\
u^{(<)}_j & \text{ if } j=\lbrace 2^{n-1-\ell}, 2^{n-1}-1\rbrace.
\end{cases}
\end{equation}
Both $u^{(>)}_j$ and $u^{(<)}_j$ can be implemented with at most $2^{\ell-1}$ controlled phase-shift gates, acting only on $\ell$ qubits. As an example of such line of reasoning, in Fig.~\ref{f_potential_barrier} is given the implementation of a potential barrier of height $u$ centered in $j=2^{n-1}$ and of range $2^{2\ell}$, corresponding to the sampling
\begin{equation}\label{eq_potential_barrier} 
u_j = \begin{cases}
u, & \text{if }2^{n-1-\ell} \leq j \leq 2^{n-1+\ell}-1,\\
0, &\text{otherwise}.
\end{cases}
\end{equation}
Moreover, a sequence of barriers of increasing height and decreasing width can be utilized to approach a $\delta$-potential. For both the former and the latter potentials, the analytic form of reflection and transmission amplitudes are well-known.
In Appendix~\ref{a_smatrix} we provide derivations in the general case and for the more specific $\delta$-potential, and also remind how to approximate a generic potential by $\delta$-pulses.

Notice that in absence of the controlled operations in Fig.~\ref{f_potential_barrier}, one would initialize a sawtooth potential spanning all the range, therefore invalidating the hypothesis of an asymptotically free particle at the beginning and end of the simulation.

As far as the gate requirement is concerned, for one iteration of the time-evolution algorithm, if $N_{f}$ and $N_{u}$ are the operations necessary to initialize the initial state and to construct $\mathcal{V}_{\delta_\tau}$, $O(\operatorname{max}(N_{f}, N_{\tau }\operatorname{max}(n^{2}, N_{u})))$ perfect gates are needed. The final state $\ket{f_\tau}$ of the simulation Eq.~\eqref{eq_time_evo_circuit} approximates $\mathcal{U}_\tau \ket{f_0}$ within $O(\delta_\tau)$, henceforth the continuous wave-function as $O(\sqrt{2^{-2n} +\delta_\tau^2})$.

\subsection{Readout of Reflection and Transmission Amplitudes}\label{s_readout}
Once the time evolution has been performed and the state $\ket{f_\tau}$ has been obtained, one has direct access to the information about reflection and transmission amplitudes. They are read out from the values of the wave-function in momentum space at asymptotic times, which are of the order $\tau_a = t_a/T$, where $t_a$ has been introduced in Sec.~\ref{s_protocol} and can be obtained by the peak value of position $\xi^{0} = x^{0} /L$ and momentum $K^{0}$ of the initial wave-function:
\begin{equation}
\tau_a = \frac{\gamma}{2\pi}\frac{\xi^{0}}{K^{0}}.
\end{equation}

\begin{figure}
\includegraphics[scale = 1.]{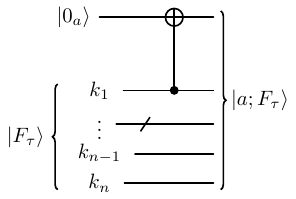}
\caption{Alignment of the particle momentum with the ancillary spin-1/2. Here $\ket{F_\tau}$ is defined in Eq.~\eqref{eq_final_state_momentum} and the final state is defined in Eq.~\eqref{eq_ancilla_particle_time_evo}.}
\label{f_readout}
\end{figure}

To obtain the final state corresponding to the wave-function in momentum space from $\ket{f_\tau}$, two operations are necessary: a qFT to map the $j$'s to the $k$'s and a phase shift for odd momenta, as in Eq.~\eqref{eq_momentum_wavefunction_k}; in other words we obtain the state $\ket{F_\tau}$ through
\begin{equation}\label{eq_final_state_momentum}
\includegraphics[scale=1]{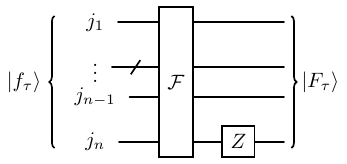}
\end{equation}
At this point, we can reproduce the ideal experiment of Sec.~\ref{s_protocol}. This is done by introducing an ancillary qubit initialized to the fiducial value $\ket{0_a}$ and, recalling that $k_1$ is associated to the sign of momentum, by applying a C-NOT operation (Fig.~\ref{f_readout}). The final state $\ket{a;F_\tau}$ can be written in the following way:
\begin{equation}\label{eq_ancilla_particle_time_evo}
\begin{split}
\ket{a;F_\tau} &= \sum_{k_2,\ldots, k_n} F_\tau^{(p\geq0)} \ket{0_{a} ;0_{k_1}}\otimes\ket{\overline{k_2\ldots k_n}}\\
&\quad + \sum_{k_2,\ldots, k_n} F_\tau^{(p<0)} \ket{1_{a} ;1_{k_1}}\otimes\ket{\overline{k_2\ldots k_n}}.
\end{split}
\end{equation}

It is evident that the ancilla and $k_1$ are aligned. We can then measure the ancilla in the computational basis to reconstruct the values of reflection and transmission amplitudes. However, since any measurement in the computational basis at the end of a circuit on only one qubit implies the measurement of all of them, we immediately realize that, as far as the readout is concerned, the ancillary qubit is redundant, since the information about scattering amplitudes is also encoded in $\ket{k_1}$. In this particular simulation, it corresponds to the sign of momentum, discriminating reflected and transmitted packets, and is by itself a spin-1/2.

More precisely, the moduli of the reflection and transmission amplitudes can be obtained by computing, after a sensible amount of iterations of the algorightm, the average value $\braket{Z_{k_1}}$, as in Eq.~\eqref{eq_spin_tomography}. As for the relative phase, one needs $\braket{X_{k_1}}$, which is read after applying the Hadamard gate on $k_1$, which maps the eigenbasis of $Z_{k_1}$ to that of $X_{k_1}$.

As suggested by Eq.~\eqref{eq_ancilla_particle_time_evo}, the time evolution will give information about the somehow ``integrated'' scattering amplitudes. However, by initializing wave-functions with reasonably small variance around the desired value of momentum, it is possible to obtain good estimates. 

\begin{figure}
\includegraphics[scale = 1.25]{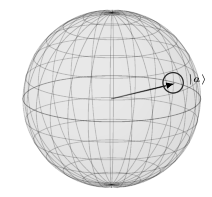}
\caption{By partitioning the sphere in uniform patches, it is possible to associate an ancilla to each individual patch, allowing for the determination of the scattering probability in a three-dimensional scattering.}
\label{f_tassellated_sphere}
\end{figure}

We conclude this section by discussing a possible generalization of this protocol to higher dimensions. It is well-known that the one-dimensional scattering is inherently different from the higher-dimensional counterpart, since in one dimension a particle can only move to the left or right, while in three dimensions it can scatter on a solid angle. It is anyway possible to reformulate the one-dimensional scattering problem in a three-dimensional form, as clearly discussed in \cite{lipkin1973quantum} and reviewed in Appendix~\ref{a_smatrix}. 
This parallelism hints at a possible generalization of our protocol to higher dimensions. While it is not the purpose of this paper to devise an optimized protocol for the three-dimensional scattering, we can point out a simple protocol which applies neatly to this case. We observe that one can couple the system to an ancillary qubit to a particular direction. Indeed, in one dimension there is only one direction and therefore only one qubit is needed. Generalizing to the three-dimesnional case, we can tessellate the sphere (Fig.~\ref{f_tassellated_sphere}) and associate to each patch an ancillary qubit. By measuring all the ancillas we can produce an histogram, containing the probability of detecting a particle within a unit of solid angle. This allows for the reconstruction of the full scattering amplitude. While this protocol may be not optimal in terms of resources, it shows the flexibility of our approach and points out to an interesting direction for future investigations.

\section{Conclusions}\label{s_conclusions}
Motivated by the current effort in computing scattering amplitudes using digital quantum simulation, in this paper we considered the scattering of a one-dimensional particle with a localized potential. We have shown how to reconstruct the values of scattering amplitudes, i.e. reflection and transmission coefficients 
$\mathcal{R}_k$ and $\mathcal{T}_k$, through the coupling of the particle with an ancillary spin-1/2. Due to the unitarity of the scattering process, we were able to successfully encode the values of reflection and transmission amplitudes in the ancillary qubit through a properly fine-tuned interaction, as discussed in Sec.~\ref{s_protocol}. It is then possible to reconstruct tomographically the state of the spin from many measurements of the polarizations, i.e. to obtain the values of reflection and transmission amplitudes. Crucially, our protocol does determine not only $\norm{\mathcal{R}_k }$ and $\norm{\mathcal{T}_k }$, but also the relative phase between the amplitudes.

We then discussed the implementation of this ideal experiment in a digital quantum simulation, by mapping the position of the particle into a register of $n$ qubits. The algorithm takes as input the initial wave-function and the potential. From the former we construct an initial state, which we map, through a Trotter-Suzuki decomposition of the Schr\"{o}dinger Hamiltonian, to a final state. The number of gates necessary to implement this time-evolution can be exponentially large for a general wave-function and potential. 
However, as discussed in Appendix~\ref{a_kinetic}, the gate requirement can be downsized by requiring that the initial wave-function to be peaked in momentum and position space and the potential to be short-ranged, as it is usually the case in quantum mechanics textbooks.

The ideal experiment we devised amounts, in the digital simulation, to the introduction of an ancillary qubit and the application of a C-NOT operation, controlled by the sign of the momentum. However, since the sign of momentum is by itself a spin-1/2, the information about reflection and transmission amplitudes can be reconstructed by the readout of such spin. In particular this can be achieved by looking at the expectation values of $\braket{Z}$ and $\braket{X}$, the latter accessible by the application of an Hadamard gate to only one qubit.

While it is well-known that in quantum mechanics only eigenvalues of Hermitian operators can be measured, there are other physical properties of a system that can be reconstructed by repeating several times the same implementation of the process. Our result exploits this general fact, i.e. we determine the scattering quantities by repeating the particle scattering a statistically significant amount of times. 

In this work, we focused on the one-dimensional case. The analysis of the protocol suggests that one can generalize it to higher dimensions, as discussed in Sec.~\ref{s_readout}. It may also be possible that the symmetries of the problem can inspire more efficient quantum algorithms to extract quantities relevant to scattering. As an example, one could study the expansion of the scattering amplitude in partial waves and couple the ancillary qubits to them rather than to the directions of the tessellated solid angle. Our study in one-dimension and its possible extensions to higher dimensions is then setting a conveninent basis for the study of scattering amplitudes in interacting theories through digital quantum simulation. 

\begin{acknowledgments}
We thank R. Konik for useful discussions.  We thank the kind hospitality and support from the programs 
"Out-of-equilibrium Dynamics and Quantum Information of Many-body Systems with Long-range Interactions" 
at KITP (Santa Barbara) and 
"Quantum Many-Body Systems Out-of-Equilibrium" at
the Institut Henri Poincaré (Paris). We acknowledge financial support from PNRR M4C2I1.3 PE\_00000023\_NQSTI Grant, Spoke A2, funded by NextGenerationEU. GM acknowledges the grant PRO3 Quantum Pathfinder.

\end{acknowledgments}

\bibliography{biblio_r_t_v2}

\appendix
\section{$S$-matrix Theory of a Localized Potential}\label{a_smatrix}

Although the $S$-matrix of the one-dimensional particle reduces to elementary linear algebra, it contains fundamental physical features, which are the cornerstones of the scattering theory in many-body quantum systems and in quantum field theory. Inspired by \cite{r_1998merzbacher_b_qm, lipkin1973quantum}, we briefly 
review the scattering properties of a particle interacting with a short-ranged potential.

As a paradigmatic example, we consider the familiar $\delta$-potential
\begin{equation}
    V(x) = g\,\delta(x-y).
\end{equation}
Its eigenfunction $\varphi_k(x)$ of energy $E = \hbar^2 k^2 /2m$ is piece-wise defined
\begin{equation}
    \varphi_k(x) = \begin{cases}
        A e^{i k x} + B e^{-i k x}, & x\leq y,\\
        C e^{i k x} + D e^{-i k x}, & x \geq y.
    \end{cases}
\end{equation}
Here $A$ ($D$) is the amplitude of the incoming particle coming from the left (right) of the potential, while $C$ ($B$) is associated to the outgoing one propagating to the right (left). As well-known, the coefficients are related by the continuity of $\varphi_k (y)$ and discontinuity of its derivative at the same point. They read
\begin{equation}
    \begin{pmatrix}
        A \\ B
    \end{pmatrix}
    = M \begin{pmatrix}
        C\\ D        
    \end{pmatrix},
\end{equation}
where the matrix
\begin{equation}\label{eq_m_matrix_delta}
    M = 
    \begin{bmatrix}
        M_{1 1} & M_{1 2}\\
        M_{2 1} & M_{2 2}
    \end{bmatrix}
    =
    \begin{bmatrix}
        1 + i\eta_{k,g} & i \eta_{k,g} e^{-i k 2y}\\
        -i \eta_{k,g} e^{i k 2y} & 1- i\eta_{k,g}
    \end{bmatrix},
\end{equation}
with $\eta_{k,g} = \frac{m}{\hbar^2}\frac{g}{k}$, is easily seen to be of 
determinant $1$ and connects the amplitudes on the left to the ones on the right of the potential.

On the other hand, the $S$-matrix connects incoming and outgoing amplitudes, and is defined -- up to permutations of the rows and columns -- as
\begin{equation}\label{eq_s_matrix_defi}
    \begin{pmatrix}
        B \\ C
    \end{pmatrix}
    = S(k) \begin{pmatrix}
        A\\ D        
    \end{pmatrix}. 
\end{equation}
The matrix $S$ is unitary ($S^{-1} = S^\dagger$) due to the conservation of the currents: $|B|^2 + |C|^2 = |A|^2 +|D|^2$. Moreover, it is also symmetric because of time-reversal for real potentials. Also, parity at $x=y$ gives an additional constraint on the diagonal entries: $S_{11} = S_{22} e^{4i k y}$. For the $\delta$-potential the $S$-matrix reads
\begin{equation}\label{eq_s_matrix_delta} 
    \quad S(k) = \begin{bmatrix}
        S_{1 1} & S_{1 2}\\
        S_{2 1} & S_{2 2}
    \end{bmatrix}
    =
    \begin{bmatrix}
        -\frac{i\eta_{k,g}}{1 + i\eta_{k,g}}e^{i k 2y} & \frac{1}{1 + i\eta_{k,g}}\\
        \frac{1}{1 + i\eta_{k,g}} & -\frac{i\eta_{k,g}}{1 + i\eta_{k,g}}e^{-i k 2y}
    \end{bmatrix}.
\end{equation}
From it we can immediately read the reflection and transmission amplitudes:
\begin{equation}\label{eq_r_t_delta_k}
    \mathcal{R}_k = S_{1 1} = \frac{M_{2 1}}{M_{1 1}}, \quad \mathcal{T}_k = S_{2 1} = \frac{1}{M_{1 1}}.
\end{equation}

In considering $\varphi_{-k}(x)$, we notice that the role of incident and escaping waves are swapped and $S(-k) = S(k)^{-1}$. Therefore
\begin{equation}\label{eq_r_t_delta_minus_k}
    \mathcal{R}_{-k} = (S^\dagger)_{1 1} = \mathcal{R}_k ^*, \quad \mathcal{T}_{-k} = (S^\dagger)_{2 1} = \mathcal{T}_k ^*,
\end{equation}
from which we obtain 
\begin{equation}
    \mathcal{R}_k \mathcal{R}_{-k}+ \mathcal{T}_k \mathcal{T}_{-k} = 1,
\end{equation}
which reduces to Eq. \eqref{eq_unitarity}, the $S$-matrix being unitary and symmetric (also equivalent to saying that the column vector $e_1 = (S_{11}, S_{21})$ is of unit norm). Notice that this result is independent of any symmetry of the potential: only conservation of probability currents and a real short-ranged potential are required. 

If we assume the potential to be symmetric at the origin, we can easily connect $\mathcal{R}_k$ and $\mathcal{T}_k$ to the phase shifts, which arise naturally in the partial wave expansion of the three-dimensional spherically symmetric scattering, see e.g. \cite{landau2013quantum}. In one dimension, if parity is a symmetry of the potential, eigenfunctions associated to energy $E = \hbar^2 k^2/2m$ can be split into even $\psi^{0}_k (x)$ and odd $\psi^{1}_k (x)$. Their asymptotic behavior reads
\begin{equation}
\begin{split}
    \psi^0_k (x) &\sim \begin{cases}
        \cos(kx - \delta_0), & x < - a/2,\\
        \cos(kx + \delta_0), & x> a/2,
    \end{cases}\\
    \psi^1_k (x) &\sim \begin{cases}
        \sin(kx - \delta_1), & x < - a/2,\\
        \sin(kx + \delta_1), & x> a/2.
    \end{cases}
    \end{split}
\end{equation}
Here, $2\delta_0$ and $2\delta_1$ are the one-dimensional analogue of the phase shifts one encounters in higher dimensions. In polar coordinates $x = r\cos\theta$, $\theta = \{ 0, \pi\}$, it is possible to rewrite the wavefunction associated to an incident wave from the left Eq.~\eqref{eq_asymptotics_eigenfunctions} as
\begin{equation}\label{eq_incoming_wave_left_mixed}
    \varphi (x) \sim e^{ikx} + g(\theta) e^{ikr}, \quad r>a/2,
\end{equation}
where the angular dependence $g(\theta)$ is related to both phase shifts and the coefficients of reflection and transmission. In one dimension it assumes only two values, since only two directions are possible in the outcome of scattering:
\begin{equation}\label{eq_phase_shift_reflection_transmission}
    \begin{split}
        g(0) &= \mathcal{T}_k -1 = \sum_{\ell = 0,1} i e^{i\delta_\ell} \sin\delta_\ell, \\
        g(\pi) & = \mathcal{R}_k = \sum_{\ell = 0,1} i (-1)^\ell e^{i\delta_\ell} \sin\delta_\ell.
    \end{split}
\end{equation}
While in one dimension spherical waves are dimensionless, in three dimensions they assume dimension of inverse length, due to their normalization:
\begin{equation}
    \varphi_{3\text{D}} \sim e^{ikx} + \frac{f(\theta)}{r} e^{ikr}.
\end{equation}
In three dimensions $f(\theta)$ is a scattering amplitude having dimensions of length. Its one-dimensional analogue is
\begin{equation}
    f(\theta) = \frac{g(\theta)}{i k} = k^{-1} \sum_{\ell = 0, 1} e^{i \ell\theta} e^{i \delta_\ell} \sin\delta_\ell,
\end{equation}
which satisfies the optical theorem
\begin{equation}
    \sum_{\theta = 0, \pi} \norm{f(\theta)} = 2 k^{-1} \operatorname{Im} f(0).
\end{equation}

Crucially, the $S$-matrix can be written in different but equivalent ways, depending on the chosen basis of wave-functions. For example, Eq.~\eqref{eq_s_matrix_defi} assumes incoming waves from the left and right as basis. To see this, it is useful to adopt a slightly modified version of Eq.~\eqref{eq_incoming_wave_left_mixed}. Generically incident waves $\psi^+$ from the left ($\theta = 0$) and right ($\theta = \pi$) can be compactly written as a combination of an incident wave $e^{-ikr}$ and an outgoing one $e^{ikr}$:
\begin{equation}
    \psi^+_{\theta '} (x) = \delta_{\theta,\, \pi-\theta'} e^{-ikr} + \left[ g(\theta' - \theta) + \delta_{\theta, \, \theta'} \right] e^{ikr}  
\end{equation}
for $r> \ell/2$. Any orthonormal combination of $\psi^+_{0}$ and $\psi^+_{\pi}$ is a valid basis. Let such basis be $\psi^+_\alpha$, with $\alpha = 1,2$:
\begin{equation}
    \psi^+_\alpha (x) = \phi_\alpha (0) \psi^+_0 (x) + \phi_\alpha(\pi) \psi^+_\pi (x),
\end{equation}
with the coefficients $\psi_\alpha(\theta)$ to be orthonormal functions. The $S$-matrix relates the amplitudes of incoming and outgoing amplitudes:
\begin{equation}
    \sum_{\theta'} \left[ g(\theta' - \theta) + \delta_{\theta, \, \theta'} \right] \phi_\alpha(\theta') = \sum_\gamma S_{\alpha\gamma} \phi_\gamma(\theta).
\end{equation}
We find the matrix elements $S_{\alpha\beta}$ by multiplying both terms of the latter equation by $\phi_\beta^*(\theta)$ and summing over $\theta$:
\begin{equation}
    S_{\alpha \beta} = \sum_\theta \sum_{\theta'} \phi^*_\beta (\theta) \left[ g(\theta' - \theta) + \delta_{\theta,\,\theta'} \right] \phi_\alpha (\theta').
\end{equation}
On the other hand, 
outgoing waves
\begin{equation}
    \psi^-_\alpha (\theta)  \equiv \sum_\gamma \psi^+_\gamma S^*_{\gamma\alpha}
\end{equation}
coincide with the solutions of the Schr\"{o}dinger equation obtained by time-reversal symmetry and reflection, which maps left to right: $\left[\psi^+_\alpha (\pi-\theta) \right]^*$. From their equality, we deduce that $S$ is symmetric: the transitions $\alpha\to\beta$ and $\beta\to\alpha$ happen with equal probabilities. 

If we specialize to the basis of incident waves $\psi^+_0$ and $\psi^+_\pi$ it is easy to see that the associated $S$ matrix has form
\begin{equation}
    S = \begin{bmatrix}
        \mathcal{T}_k  & \mathcal{R}_k\\
        \mathcal{R}_k & \mathcal{T}_k
    \end{bmatrix},
\end{equation}
coincident to Eq.~\eqref{eq_s_matrix_defi} up to permutations of rows and columns. On the other hand if we write even and odd wave-functions in terms of incoming and outgoing waves
\begin{equation}
    \psi_\ell (x) = - \left[ e^{i\ell(\pi-\theta)} e^{-ikr} + e^{2i\delta_\ell} e^{i\ell\theta} e^{ikr}\right],
\end{equation}
it is immediate to see that the $S$-matrix is diagonal and its entries are
\begin{equation}
    S_{\ell\ell'} = e^{2i\delta_\ell} \delta_{\ell, \, \ell'}.
\end{equation}
The latter equation closes the circle. In one dimension the even and odd phase shifts are the eigenvalues of the $S$-matrix and can be extracted from the reflection and transmission coefficients through Eq.~\eqref{eq_phase_shift_reflection_transmission}.

As an example, we can compute the phase shifts for the $\delta$-potential centered at the origin:
\begin{equation}
 \tan \delta_0 = - \eta_{k,g} , \quad \delta_1 = 0.
\end{equation}
Indeed, the phase shift associated with the odd solution must vanish because of the continuity at the origin. This means that for a $\delta$-potential, only the component of a wave coming from the even solution contributes to the scattering. A consequence of this is that $f(\theta)$ is independent of $\theta$, since
\begin{equation}
    g(\theta) = \mathcal{T}_k -1 = \mathcal{R}_k = \frac{1}{2} (e^{2i\delta_0} - 1) = \frac{1-i\eta_{k,g}}{1+i\eta_{k,g}}.
\end{equation}

\subsection{Approximation of a Generic Potential with $\delta$-Pulses}
\begin{figure}
    \centering
    \includegraphics[scale = 0.35]{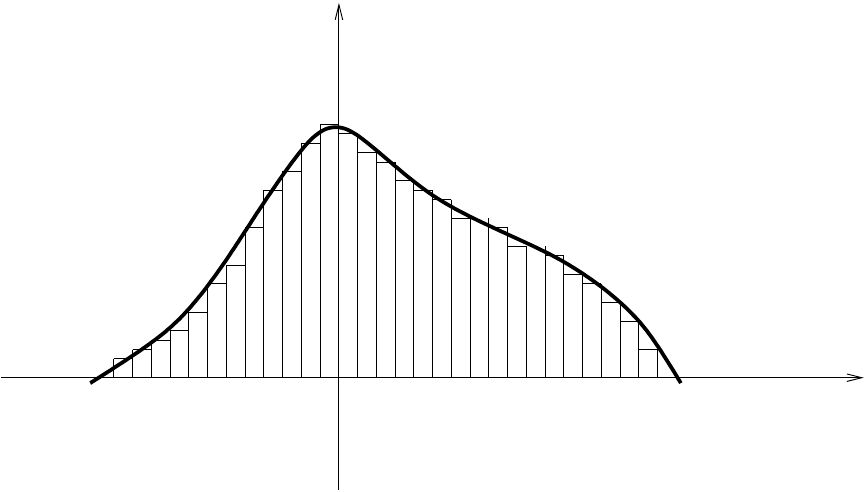}
    \caption{Approximation of a generic short-ranged potential through piecewise constant barriers.}
    \label{f_general_potential}
\end{figure}

Any localized potential in the range $-a/2 \leq x\leq a/2$ can be approximated by piece-wise constant functions, see e.g. Fig. \ref{f_general_potential}. 
However, if we sample a sufficiently large number of points $N$ of the potential, a good approximation is also given by a sequence of pulses
\begin{equation}\label{eq_approx_potential}
    V_N (x) = \sum_{j=0}^{N-1} V_j \frac{a}{N} \delta\left(x+\frac{a}{2} - \frac{j}{N}a\right),
\end{equation}
with $V_j = V\left(-\frac{a}{2} + \frac{j}{N}a\right)$.

For this potential, the amplitudes of the waves from the left $(x\leq -a/2)$ and from the right $(x\geq a/2)$ are easily connected. First of all, notice that for $\delta$-potential considered previously we can rewrite 
\begin{equation}
    \begin{pmatrix}
        A e^{i k y} \\ B e^{-i k y}
    \end{pmatrix}
    = M_0 \begin{pmatrix}
        Ce^{i k y}\\ De^{-i k y}       
    \end{pmatrix},
\end{equation}
\begin{equation}
M_0 =  \begin{bmatrix}
        1 + i\eta_{k,g} & i \eta_{k,g}\\
        -i \eta_{k,g}  & 1- i\eta_{k,g}
    \end{bmatrix},
\end{equation}
where $M_0$ is the matrix associated to the same potential, however located at the origin. Hence each time the particle reaches $x_j = -a/2 + j a/N$, a matrix $M_0$ with parameter
\begin{equation}
    \eta_{k,j} = \frac{2m}{\hbar^2}\frac{a}{N}\frac{V_j}{2k}
\end{equation}
connects the amplitudes. Then, the incoming wave travels a distance $\Delta x = a/N$ and again a scattering with matrix $M_0$ occurs. Defining the matrices
\begin{align}
     M_0^{(j)} &=  \begin{bmatrix}
        1 + i\eta_{k,j} & i \eta_{k,j}\\
        -i \eta_{k,j}  & 1- i\eta_{k,j}
    \end{bmatrix}, \\ 
    D &= \begin{bmatrix}
        e^{-ik a/N} & 0 \\
    0 & e^{ik a/N}
    \end{bmatrix}, \\
    U &= \begin{bmatrix}
        e^{ik a/2} & 0 \\
    0 & e^{-ik a/2}
    \end{bmatrix},
\end{align}
allows us writing
\begin{equation}
    \begin{pmatrix}
        A \\ B
    \end{pmatrix}
    = \mathcal{M}_N \begin{pmatrix}
        C\\ D        
    \end{pmatrix}, 
\end{equation}
\begin{equation}
\mathcal{M}_N = U \left[\prod_{j=0}^{N-1} M_0^{(j)} D\right] U.
\end{equation}
The reflection and transmission coefficients can be easily read from the elements of $\mathcal{M}_N$, as in Eq. \eqref{eq_r_t_delta_k}. 

Despite the simplicity of this method, the resummation of the single matrix elements is nontrivial. Even for the prototypical case of the potential barrier of height $V_0$, the amplitudes for $E<V_0$ contain nontrivial hyperbolic functions, which are purely due to tunneling in the classically prohibited region. The matrix $\mathcal{M}$ reads
\begin{widetext}
\begin{equation}
\mathcal{M} = \begin{bmatrix}
\left(\cosh  \kappa a+\frac{i}{2}\varepsilon_{-} \sinh  \kappa a\right) e^{ i k a} & \frac{i }{2}\varepsilon_{+} \sinh  \kappa a \\
-\frac{i }{2}\varepsilon_{+} \sinh  \kappa a & \left(\cosh  \kappa a-\frac{i }{2}\varepsilon_{-} \sinh  \kappa a\right) e^{- i k a}
\end{bmatrix},
\end{equation}
\end{widetext}
with $\kappa = \sqrt{2m(V_0-E)/ \hbar^2}$ and $ \varepsilon_{\pm} = \frac{\kappa}{k}\pm \frac{k}{\kappa}$.

\begin{figure*}
\centering
\begin{subfigure}{0.45\textwidth}  
\includegraphics[scale=0.8]{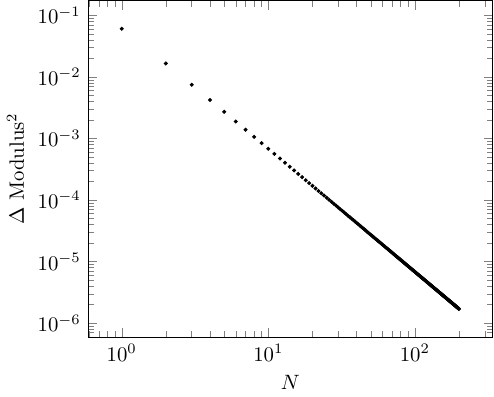}
\caption{}
\end{subfigure}
\begin{subfigure}{0.45\textwidth}  
\includegraphics[scale=0.8]{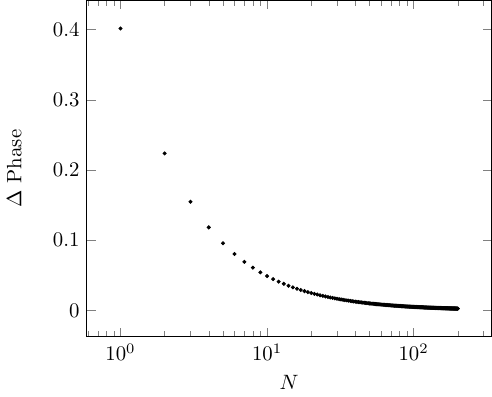}
\caption{}
\end{subfigure}
\caption{Plots of the approximation error of modulus (a) and phase (b) for the reflection coefficient of a potential barrier of opacity $ \sqrt{2m V_0/\hbar^2} a =1$ and energy $E/V_0 = 1/\sqrt{2}$ for $N\leq 200$.}
\label{f_barrier}
\end{figure*}

This feature is completely absent in presence of finitely many $\delta$-potentials, since in every region of space the eigenfunction is a superposition of plane waves and can only be retrieved in the limit of infinite $N$. However, as can be seen from Fig. \ref{f_barrier}, the convergence is incredibly fast both in the modulus and the phase, even for a sampling of a few hundred points.

\begin{figure*}
\centering
\begin{subfigure}{0.3\textwidth}  
\includegraphics[scale=0.6]{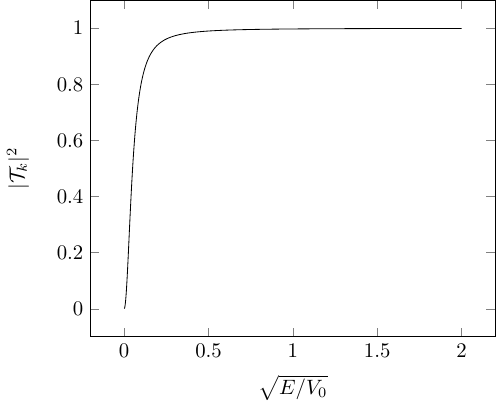}
\caption{}
\end{subfigure}\hfill
\begin{subfigure}{0.3\textwidth}  
\includegraphics[scale=0.6]{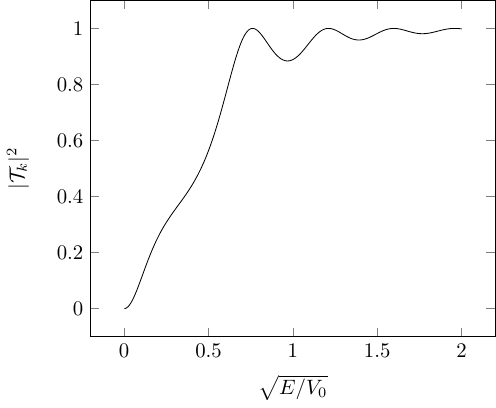}
\caption{}
\end{subfigure}\hfill
\begin{subfigure}{0.3\textwidth}  
\includegraphics[scale=0.6]{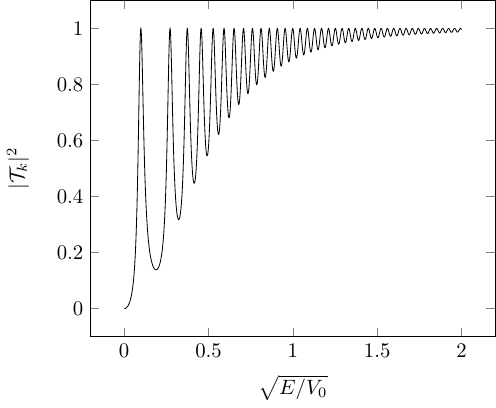}
\caption{}
\end{subfigure}
\caption{Plots of the transmission probability for various values of $\sqrt{2m V_0/\hbar^2} a $ ($0.1$, $10$ and $100$ respectively for (a), (b) and (c)) for an attractive potential barrier, using the sampling method with $N=1000$.}
\label{f_t_resonance}
\end{figure*}

Our method can also be applied to attractive potentials, for the $M_0$ matrix holds for any sign of the coupling. It is then possible to efficiently evaluate the transmission probability $|\mathcal{T}_k|^2$, which presents resonances. For a potential barrier of height $-V_0$, a simple analytic continuation of the matrix $\mathcal{M}$ shows that resonances ($|\mathcal{T}_k|^2=1$) are to be found at values of $\sqrt{2m (V_0+E)/\hbar^2}a = \ell\pi$ for integer $\ell$ 
and minima at $\sqrt{2m (V_0+E)/\hbar^2}a = \ell\frac{\pi}{2}$ 
(Fig.~\ref{f_t_resonance}).

\section{Digital Implementation of the Time-Evolution Operators}\label{a_kinetic}
In this appendix, we discuss some specific constructions for the phase shift Eq.~\eqref{eq_unitary_phase_shift} of Sec.~\ref{s_simulation} and show that there are indeed some cases in which an efficient implementation exists, for certain kind of functions.

\begin{figure}
\includegraphics[scale=0.75]{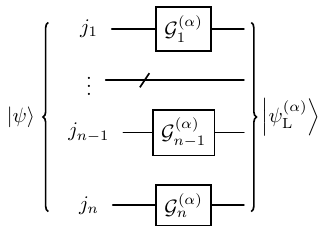}
\caption{Realization of the linear phase-shift Eq.~\eqref{eq_linear_phase_shift}. Here $\mathcal{G}^{(\alpha)}_{\ell}$ is defined in Eq.~\eqref{eq_building_block_linear}.}
\label{f_linear_phase_shift}
\end{figure}

As a first example, we consider the linear phase shift $g^{(\alpha)}_{\text{L}}(j) = \alpha j$, for real $\alpha$:
\begin{equation}\label{eq_linear_phase_shift}
\ket{\psi}  = \sum_{j=0}^{2^n-1} c_j \ket{j} \to \ket{\psi^{(\alpha)}_{\text{L}}}=\sum_{j=0}^{2^n-1} c_j e^{i\alpha j} \ket{j}.
\end{equation} 
The construction is immediate by considering that in the convention Eq.~\eqref{eq_convention_bits} we may express, as remarked in \cite{r_2008benenti_qc_oneparticle_scattering},
\begin{equation}
e^{i \alpha j} = \prod_{\ell=1}^{n}e^{i \alpha j_\ell 2^{n-\ell}}.
\end{equation}
As such the circuit implementing the linear phase shift is given in terms of the building-block matrix
\begin{equation}\label{eq_building_block_linear}
\mathcal{G}^{(\alpha)}_\ell = \begin{bmatrix}
1 & 0 \\
0 & e^{i \alpha 2^{n-\ell}}
\end{bmatrix}
\end{equation}
and shown in Fig.~\ref{f_linear_phase_shift}. Indeed, this particular transformation requires $O(n)$ gates, without controlled operations.

\begin{figure}[t]
\includegraphics[scale=0.6]{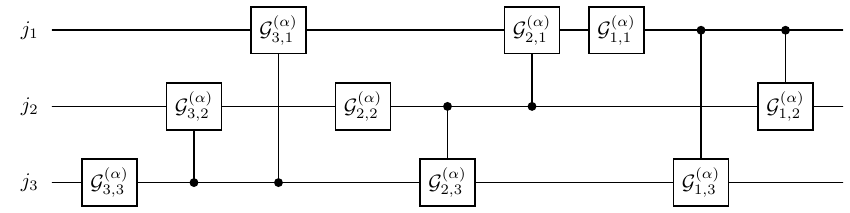}
\caption{Realization of the quadratic phase-shift Eq.~\eqref{eq_quadtratic_phase_shift} for the case for $n=3$ qubits. Here $\mathcal{G}^{(\alpha)}_{\ell_1, \ell_2}$ is defined in Eq.~\eqref{eq_quadratic_gates}.}
\label{f_quadratic_phase_3qubits}
\end{figure}

Allowing for the use of controlled operations, one can construct the operator implementing a quadratic phase-shift
$g^{(\alpha)}_{\text{Q}}(j) = \alpha j^2$, i.e. the unitary transformation
\begin{equation}\label{eq_quadtratic_phase_shift}
\ket{\psi}  = \sum_{j=0}^{2^n-1} c_j \ket{j} \to \ket{\psi^{(\alpha)}_{\text{Q}}}=\sum_{j=0}^{2^n-1} c_j e^{i\alpha j^2} \ket{j}.
\end{equation}
The implementation with gates follows directly from the identity
\begin{equation}
e^{i\alpha j^2} = \prod_{\ell_1 =1}^{n}\prod_{\ell_2 =1}^{n}e^{i\alpha j_{\ell_1} j_{\ell_2} 2^{2n-\ell_1-\ell_2}}. 
\end{equation}
In principle $n^2$ gates of the form
\begin{equation}\label{eq_quadratic_gates}
\mathcal{G}^{(\alpha)}_{\ell_1, \ell_2} = \begin{bmatrix}
1 & 0 \\
0 & e^{i \alpha 2^{2n-(\ell_1+\ell_2)}}
\end{bmatrix}.
\end{equation}
are sufficient: $n$ unitaries of the type $\mathcal{G}^{(\alpha)}_{\ell, \ell}$ and $(n^2-n)$ controlled operations of the type $\mathcal{G}^{(\alpha)}_{\ell_1, \ell_2}$. In Fig.~\ref{f_quadratic_phase_3qubits} we present a realization of the quadratic phase-shift for the case for $n=3$ qubits, but the generalization to arbitrary $n$ is straightforward.

The implementation of generic powers $g(j)= j^p$ follows directly from the quadratic case, if one wants to use controlled operations with $(p-1)$ control qubits and the building-block unitary
\begin{equation}
\mathcal{G}^{(\alpha)}_{\ell_1, \ldots, \ell_p} = \begin{bmatrix}
1 & 0 \\
0 & e^{i \alpha 2^{pn-(\ell_1+\ldots+\ell_p)}}
\end{bmatrix}.
\end{equation}
\begin{figure}
\includegraphics[scale=0.75]{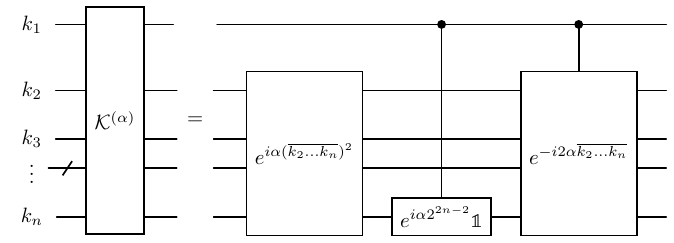}
\caption{Implementation of the kinetic operator Eq.~\eqref{eq_kinetic_unitary_transformation}, assuming that linear (Eq.~\eqref{eq_linear_phase_shift}) and quadratic (Eq.~\eqref{eq_quadtratic_phase_shift}) phase shifts are used as subroutines. The convention for the representation of bits is given in Eq.~\eqref{eq_convention_k}.}
\label{f_kinetic}
\end{figure}

The linear phase shift $g^{(\alpha)}_{\text{L}}(j)$ and the quadratic one $g^{(\alpha)}_{\text{Q}}(j)$ are the building blocks to implement the kinetic operator $\mathcal{K}_{\delta_\tau}$ of Sec.~\ref{s_simulation}. The general structure of such operator is, because of Eq.~\eqref{eq_momentum_k},
\begin{equation}\label{eq_kinetic_unitary_transformation}
\widehat{\mathcal{K}}^{(\alpha)} \ket{k} = \begin{cases}
e^{i\alpha k^2}\ket{k}, & \text{ if } k=\lbrace 0, \ldots, 2^{n-1}-1\rbrace,\\
e^{i \alpha (2^n - k)^2}\ket{k}, & \text{ if } k=\lbrace 2^{n-1}, \ldots, 2^n -1\rbrace.
\end{cases}
\end{equation}
We recover $\mathcal{K}_{\delta_\tau}$ when $\alpha = - \gamma (2\pi)^2 \delta_\tau$. Moreover, by expanding the square in the latter equation, it is immediate to assign the role of a control to the bit $k_1$; indeed
\begin{equation}
\widehat{\mathcal{K}}^{(\alpha)} \ket{k} = e^{i\alpha \left(\overline{k_2 \ldots k_n}\right)^2} e^{\delta_{k_1, 1} \left[-i\alpha \left(2\overline{k_2 \ldots k_n}-2^{2n-2}\right) \right]}\ket{k}.
\end{equation}
Hence $\mathcal{K}_\tau$ can be implemented with $n^2 - n + 2$ gates (either single-qubit phase shifts or controlled with only one control), i.e. is still $O(n^2)$. A circuit representation of the kinetic gate is given in Fig.~\ref{f_kinetic}. 

\end{document}